\documentclass[a4paper,aps,pre,twocolumn,groupedaddress,showkeys,showpacs,floats,floatats,floatfix]{revtex4-1}
\usepackage[utf8]{inputenc}
\usepackage[english]{babel}
\usepackage{graphicx}
\usepackage{dcolumn} 
\usepackage{bm}
\usepackage{natbib}
\usepackage{latexsym}
\usepackage{mathrsfs}
\usepackage{amssymb}
\usepackage{amsmath}
\usepackage{amscd}
\usepackage[usenames,dvipsnames]{xcolor}
\usepackage{pifont}
\usepackage{pstricks,pst-node,pst-text,pst-3d}
\usepackage{verbatim}
\usepackage{ulem}
\newrgbcolor{redcolor}{1.0 0.0 0.0}

\newrgbcolor{bluecolor}{0.0 0.0 1.0}

\begin{document}

\title{Characterizing Weak Chaos using Time Series of Lyapunov
  Exponents} 
\author{R.M. da Silva$^{1}$, C. Manchein$^{1}$, M.W. Beims$^{2,3}$,
  E.G. Altmann$^{3}$} 
\affiliation{$^1$Departamento de F\'\i sica, Universidade do Estado de
  Santa Catarina, 89219-710 Joinville, Brazil}  
\affiliation{$^2$Departamento de F\'\i sica, Universidade Federal do
  Paran\'a, 81531-980 Curitiba, Brazil} 
\affiliation{$^3$Max-Planck-Institute for the Physics of Complex
  Systems, N\"othnitzer Str.~38, 01187, Dresden, Germany, EU} 
\date{\today}
%
\begin{abstract}
We investigate chaos in mixed-phase-space Hamiltonian systems using
time series of the finite-time Lyapunov exponents. The methodology we
propose uses the  number of Lyapunov exponents close to zero to define
regimes of ordered (stickiness), semi-ordered (or semi-chaotic), and
strongly chaotic  motion. The dynamics is then investigated looking at
the consecutive time spent in each regime, the transition between
different regimes, and the regions in the phase-space associated to
them. Applying our methodology to a chain of  coupled standard maps we
obtain: (i) {that} it allows for an improved 
numerical characterization of stickiness in high-dimensional Hamiltonian 
systems, when compared to the previous analyses based on the distribution 
of recurrence times; (ii) {that the} transition  probabilities 
between different regimes are determined by the phase-space volume associated 
to the corresponding regions; (iii) the 
dependence of the Lyapunov exponents with {the coupling strength.}
\end{abstract}

\maketitle

\section{Introduction}
\label{intro}
In weakly chaotic Hamiltonian systems regions of regular (periodic and 
quasi-periodic) and chaotic motion typically coexist in the 
phase-space~\cite{Lichtenberg,Meiss}. In high dimensions, due to
Arnold diffusion, all initial conditions leading to chaotic motion are
connected in the phase-space building a single chaotic
component~\cite{Lichtenberg}. Even if the volume of the regular
regions becomes vanishingly small, as expected for high-dimensional
nonlinear systems, the dynamics inside the chaotic component of the
phase-space is strongly affected by such regions. This happens because 
trajectories approaching non-hyperbolic  regions or regular motion 
remain a long time close to them
before visiting again other parts of the chaotic component of the
phase-space. This signature of weak mixing (or weak chaos) is known as
stickiness~\cite{Chir-Shep,Artuso,ZaslavskyPhysicsReports,AltmannThesis,
Manchein,Cristadoro}.      

Since Chirikov-Shepelyansky~\cite{Chir-Shep}, the main quantification
of stickiness in Hamiltonian systems has been through the fat-tail
distribution of Poincar\'e recurrence times (see, \textit{e.g.},
~\cite{Cristadoro,Artuso,AltmannKantz}). {An alternative approach is 
to use finite-time Lyapunov exponents 
(FTLEs)~\cite{GrassbergerKantz,Viana,Harle}, with  
recent applications
using large deviation techniques~\cite{Manchein,New} and the 
cumulants~\cite{cesar12,cesar14} of
the FTLE distribution.} In area-preserving maps,
stickiness generically occurs 
at the border of $2$-dimensional Kolmogorov-Arnold-Moser (KAM)
island~\cite{Lichtenberg} ({\it i.e.}, at $1$-dimensional tori). 
The recurrence time is a measure of the time the trajectory spends around 
such structures before returning to the chaotic sea 
(stickiness happens also to one-parameter families of parabolic orbits 
\cite{gd95,amk06} and even to isolated parabolic fixed points ~\cite{Artuso,
Artuso2,sma14}). 
{Near the non-hyperbolic structures,  the local instability of chaotic 
trajectories is reduced so that FTLEs  can be used to characterize phase-space
regions of interest~\cite{Viana,Harle,New,Contopoulos,Malagoli}}.
Stickiness has been studied also in higher-dimensional
systems \cite{AltmannKantz,cesar12,cesar14,bountis,Malagoli,GrassbergerKantz},
long recurrence times can be due to different non-hyperbolic regions and 
tori of different dimensionalities \cite{lange10}. An improved 
characterization of 
stickiness events (long recurrence time) requires thus to measure the number 
of stable and unstable directions in the trajectory during this event. 
Froeschl\'e conjectured that lower-dimensional tori could not 
exist~\cite{Froeschle1,Froeschle2,Lichtenberg}. In early studies in the 
80's such events of stickiness to lower dimensional tori were reported in 
some systems~\cite{Malagoli} but were not found in other 
examples~\cite{GrassbergerKantz}. Even if invariant tori do not exist, 
small local Lyapunov exponent could effectively act as a lower-dimensional 
trap. This is similar to almost invariant sets \cite{dellnitz97,froyland09}, 
which are regions in phase-space where typical trajectories stay (on average) 
for long periods of time. 

In this paper we introduce a methodology that uses time-series of local 
Lyapunov exponents to define regimes of ordered, semi-ordered and totally 
chaotic motion and obtain an improved characterization of stickiness in 
high-dimensional Hamiltonian systems. We illustrate this general procedure in 
a chain of coupled standard maps and confirm that stickiness events of 
different times length are dominated by trajectories with different
FTLEs. {A significant  
improvement of the characterization of sticky motion in high-dimensional 
systems is found.} We also characterize the 
FTLEs for small couplings and compare them to 
expected universal properties in fully chaotic systems~\cite{Daido1}. The 
method proposed here is general and can be used to investigate Hamiltonian 
systems in any dimension. 

The paper is divided as follows. In Sec.~\ref{mod} we describe the
Hamiltonian model we use to illustrate our method. In
Sec.~\ref{methods} we introduce our method to compute and analyze time
series of local Lyapunov exponents. This methodology is then applied
to the symplectic model of coupled standard maps in
Sec.~\ref{nr}. Section~\ref{cc} summarizes the main results of 
 the paper. 

\section{The coupled maps model}
\label{mod}
We use a time-discrete $2N$-dimensional Hamiltonian system obtained as
the composition $\mathbf{T} \circ \mathbf{M}$ of independent one-step
iteration of $N$ symplectic $2$-dimensional maps
$\mathbf{M}=(M_1,...,M_N)$ and a symplectic coupling $\mathbf{T} =
(T_1,...,T_N)$. As a representative example of $2$-dimensional maps we
choose for our numerical investigation the standard map: 
\renewcommand{\arraystretch}{1.2}
\begin{equation}
  \label{mp-acop1}
  \mathbf{M_i}\left(
  \begin{array}{c}
    p_i \\
    x_i \\
  \end{array}
  \right) = \left(
  \begin{array}{llll}
    p_i + K_i \sin(2\pi x_i) & \hspace{0.1cm} \mathrm{mod} \hspace{0.2cm} 1 \\
    x_i + p_i + K_i \sin(2\pi x_i) & \hspace{0.1cm} \mathrm{mod} \hspace{0.2cm} 1 \\
  \end{array}
  \right),
\end{equation}

\noindent and for the coupling 

\begin{equation}
  \label{mp-acop2}
  \mathbf{T_i} \left(
  \begin{array}{c}
    p_i \\
    x_i \\
  \end{array}
  \right) = \left(
  \begin{array}{llll}
    p_i + \sum_{j=1}^{N} \xi_{i,j} \hspace{0.05cm} \sin[2\pi (x_i - x_j)] \\
    x_i \\
  \end{array}
  \right),
\end{equation}

\renewcommand{\arraystretch}{1}
\noindent with $\xi_{i,j} = \xi_{j,i} = \frac{\xi}{\sqrt{N-1}}$ (all-to-all 
coupling). The motivation for working with this system is that in the
limit of small coupling $\xi\rightarrow 0$ it can be understood
looking at the dynamics of the $N$ uncoupled maps. This system was
studied in Refs.~\cite{AltmannKantz,AltmannThesis} using recurrence
time distribution.  This allow us to critically compare the benefits
of our methodology. In all numerical simulations we used $K_1=0.5214$
for the map $M_1$ and $K_2=K_3=0.5108$ for the maps $M_2$ and $M_3$.   

\section{Method}
\label{methods}
In this section we describe the method proposed in this work. To be
illustrative, we present numerical simulations for the system defined
in Sec.~\ref{mod}. 

\subsection{Lyapunov spectrum and the classification of ordered,
  semi-ordered or semi-chaotic,  and chaotic regimes}     
\label{regions}

Consider a chaotic trajectory in a closed Hamiltonian system which, after 
reducing the phase-space dimension due to global invariant of motion, has $N$
degree of freedoms. For long times $t$ the trajectory ergodically fills the 
whole chaotic component of the phase-space which is characterized by a spectrum 
of $N$ Lyapunov exponents $\{\lambda_{i=1\ldots N}^{(\infty)}\}$, where 
$\lambda^{(\infty)}_1 > \lambda^{(\infty)}_2,\ldots,\lambda^{(\infty)}_N > 0$ 
\footnote{Without loss of generality we focus on the $N$ largest Lyapunov 
exponents because due to the symplectic character of Hamiltonian systems the 
others $N$ exponents are simply $\lambda_{N+1}=-\lambda_N, \lambda_{N+2}=
-\lambda_{N-1}, \ldots \lambda_{2N}=-\lambda_1$.}. The central ingredient of 
our analysis is the spectrum of FTLEs computed along a trajectory during a 
window of size $\omega$ where we obtain a time dependent spectrum 
$\{\lambda^{(\omega)}_i\}(t)=\{\lambda_i^{(\omega)}\}$. The window size 
$\omega$ has to be sufficiently small to guarantee a good resolution of 
the temporal variation of the $\lambda^{(\omega)}_i$'s, but sufficiently large 
in order to have a reliable estimation (see Refs.~\cite{GrassbergerKantz,Viana,
Harle}). The probability density function of $\lambda_i^{(\omega)}$ has been 
extensively studied~\cite{GrassbergerKantz,Viana,Harle,Manchein}. Here we go 
beyond the study of the probability density function and explore temporal 
properties in the time series of $\{\lambda_i^{(\omega)}\}$. 

\begin{figure*}[!htb]
  \centering
  \includegraphics*[width=0.9\columnwidth]{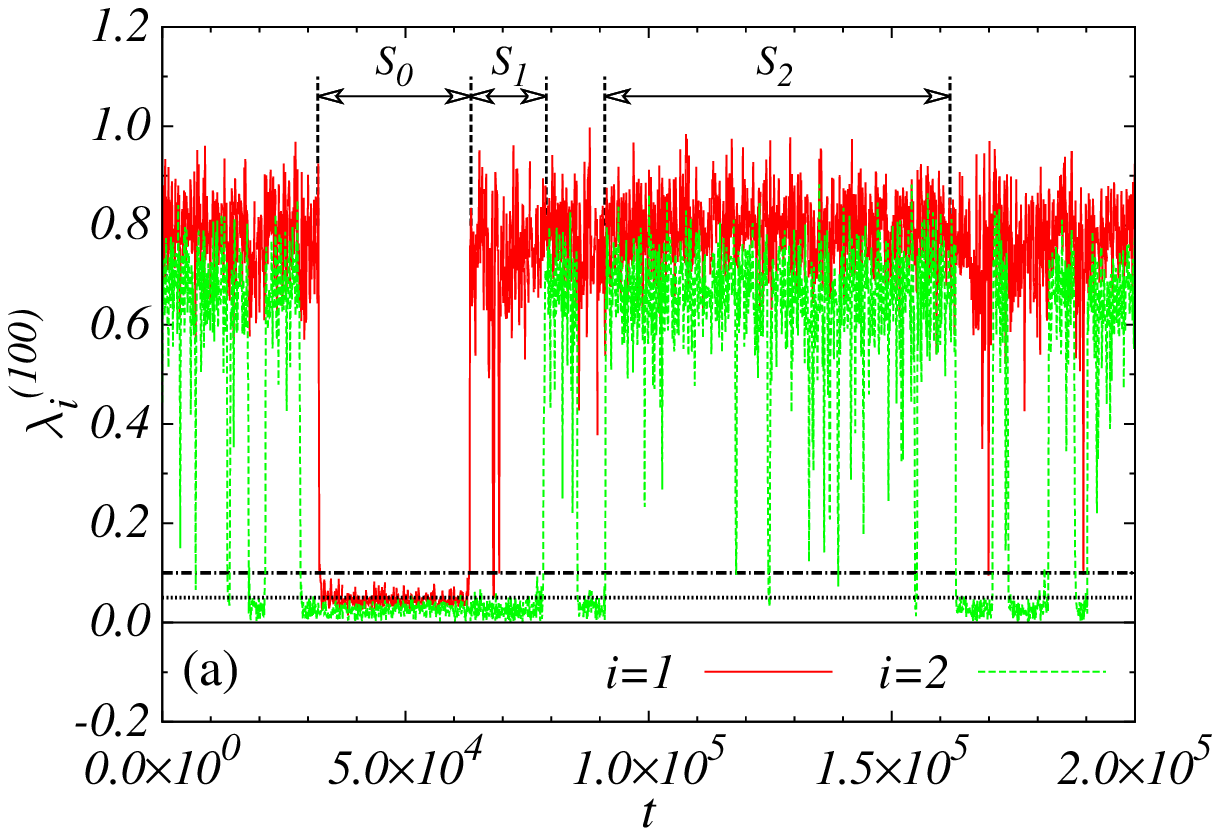}
  \includegraphics*[width=0.9\columnwidth]{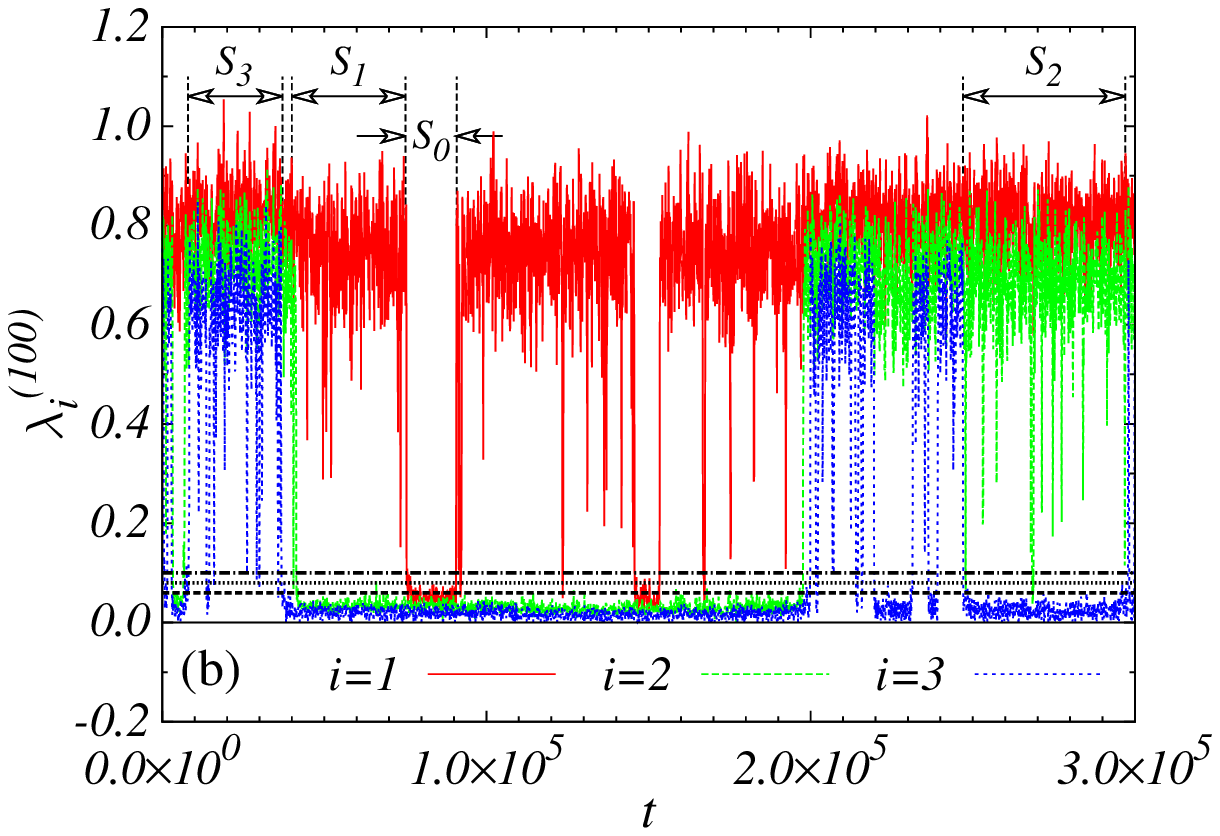}
  \caption{(Color online) Illustration of the method proposed to
    define the regimes $S_M$. Time series of the spectrum of FTLEs
    {$\{\lambda_i^{(\omega=100)}\}$, with $i=1,\ldots, N$}, for the
    map~(\ref{mp-acop1})-(\ref{mp-acop2}) {with $\xi=10^{-3}$}. 
    Panel (a): Case $N=2$
    and the thresholds $\varepsilon_1=0.1$ and $\varepsilon_2=0.05$
    are represented by dash-dotted and dotted lines
    respectively. Panel (b): Case $N=3$ and the thresholds
    $\varepsilon_1=0.1$, $\varepsilon_2=0.08$ and
    $\varepsilon_3=0.06$ are represented by dash-dotted, dotted and
    dashed lines respectively.}  
  \label{lyapsloct}
\end{figure*}

Figures \ref{lyapsloct}(a) and (b), for $N=2$ and $3$
  respectively, show the time series of $\lambda^{(\omega)}_i$,
  ($i=1,\ldots,N$). The sharp 
transitions towards $\lambda_i^{(\omega)} \approx 0$ motivates the
classification in regimes of motion~\cite{Contopoulos,Malagoli} as (a)
{\it ordered} ($\lambda^{(\omega)}_{1,2} \approx 0$); (b) {\it
  semi-ordered  or semi-chaotic} ($\lambda^{(\omega)}_1>0;
\lambda^{(\omega)}_2\approx0$); and (c) {\it chaotic}
($\lambda^{(\omega)}_{1,2}>0$). For a system with $N$ degrees of
freedom we will say that the trajectory is in a regime of type
$S_M^{(N)}$ if it has $M$ local Lyapunov exponents
$\lambda^{(\omega)}_i > \varepsilon_i$, where {$\varepsilon_i \ll
\lambda_i^{(\infty)}$} are the small thresholds. This means that
  $S_0^{(N)}$ and  $S_N^{(N)}$ are ordered and chaotic regimes
  respectively. Whenever there is no ambiguity, we will drop the
superscript $S_M^{(N)}=S_M$ to have a simpler notation.
 
Practical implementations of the general method described above require the 
choice of a few parameters and conventions.  First of all, the window size 
$\omega$ and the threshold $\varepsilon_i$ 
{directly affect the classification in
regimes}. They can be thought as  the phase-space resolution 
of the analysis and should be chosen so that it provides maximal information 
about the regions of interest. Unless stated otherwise, we use $\omega=100$ and 
$\varepsilon_i \approx 0.10\langle \lambda_i^{(\omega)} \rangle$, where 
$\langle \ldots \rangle$ denotes average over $t$, where $t=1,\ldots,t_L$ 
({even though the {\it classification} in regimes is strongly $\omega$-dependent,
  our {\it conclusions} are not sensitively affected by variations
around the chosen values)}. 
Another 
important choice is the method for computation of the FTLEs. We use Benettin’s 
algorithm \cite{bggs80,wolf85}, which includes the Gram–Schmidt 
re-orthonormalization procedure. The decreasing order of
{$\lambda_i^{(\omega)}$} is valid on average, but inversions of
the order ({$\lambda_{i+1}^{(\omega)} > \lambda_i^{(\omega)}$}) may
happen  for some times $t$ and we have chosen to impose the order of
{$\lambda_i^{(\omega)}$} for all $t$. Finally, it is possible to
decide how to sample the time series {$\lambda_i^{(\omega)}$}.  
While the FTLEs are defined for all $t$, there is a trivial
correlation between the values of {FTLEs}  
inside a window of size $\omega$ because they are computed using the same points 
of the trajectory. In order to avoid this trivial correlation the series of 
{$\lambda_i^{(\omega)}$} can be computed using non-overlapping
windows, {\it i.e.}  plotting {$\lambda_i^{(\omega)}$} only every
$\omega$ time steps (a choice we adopt in our simulations).   

\subsection{Identifying phase-space regions}
\label{ph}

In order to understand the properties of the time series 
$\{\lambda^{(\omega)}_i\}$ it is useful to consider the phase-space regions 
associated to each regime $S_M$. We denote by $\mu(A)$ the phase-space volume 
(Liouville measure) of region $A$ in the bounded phase-space~$\Gamma$, {\it i.e.} 
$\mu(\Gamma)\equiv1$. The most important distinction is between the regions of 
regular $\Gamma_{\text{regular}}$ and chaotic $\Gamma_{\text{chaos}}$ motion. 
In Hamiltonian systems, typically $\mu(\Gamma_{\text{chaos}})>0$ and 
$\mu(\Gamma_{\text{regular}})>0$. In principle, the regular region 
$\Gamma_\text{regular}$ can be subdivided according to the dimensionality of 
the tori. However, according to Froeschl\'e's conjecture, in a $2N$-dimensional 
phase-space, tori with dimension $N$ have positive measure and thus 
$\mu(\Gamma_{\text{regular}})=\mu(\Gamma_{\text{tori}})$~\cite{Froeschle1,
Froeschle2,Lichtenberg}. For $N>1$, the chaotic region $\Gamma_{\text{chaos}}$ 
is expected to build a single ergodic component because tori of $N$ dimension 
do not partition the $2N$-dimensional phase-space in different regions and 
therefore any chaotic trajectories eventually explores (through Arnold Diffusion) 
the whole $\Gamma_{\text{chaos}}$. Our interest is not to test the Froeschl\'e 
conjecture or Arnold diffusion, but to show the insights about the chaotic 
dynamics we can obtain using the time series of $\{\lambda^{(\omega)}_i\}$ 
together with the definition of the regimes $S_M$. One application is to use 
the regimes $S_M$ to split the chaotic component of the phase-space in 
meaningful components. This is done by considering the set of points 
$\boldsymbol{X}_M^{(N)}$ in the phase-space leading to each regime $S_M$ as 

\begin{equation}
  \boldsymbol{X}_M^{(N)}=\lim_{t_L \rightarrow \infty}
  \boldsymbol{x}_t(\boldsymbol{x}_t \in S_M), 
\end{equation} 
where $t_L$ is the total length of the trajectory and {$\boldsymbol{x}_t \in 
S_M$} indicates that at time $t$ the trajectory at {$\boldsymbol{x}_t$} had 
$\{\lambda_i^{(\omega)}\} \in S_M$. 

\begin{figure*}[!t]
  \centering
  \includegraphics*[width=1.7\columnwidth]{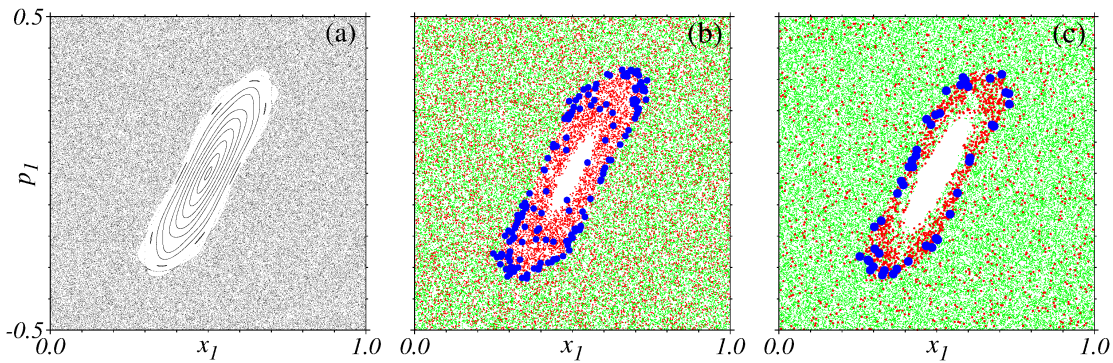}
  \caption{(Color online) Phase-space projected in $(x_1,p_1)$
    for different configuration of the $N$-coupled standard maps
    defined in Sec.~\ref{mod}. (a) $N=1$ (uncoupled case), showing
    $10^2$ randomly started initial conditions and plotting as dots
    $10^4$ iterations of each of them. A large KAM island can be seen
    at the center of the plot; (b) $N=2$ and coupling strength $\xi =
    10^{-3}$; (c) $N=3$ and $\xi = 10^{-3}$. Symbols with different
    colors in (b,c) show points $\boldsymbol{x}_{t}\in S_M$ belonging
    to regimes $S_0$ (blue circles), $S_1$ (red points), and $S_2$
    (green points). These points were computed starting a single
    trajectory in the chaotic region of all maps and iterating it 
    $5 \times  10^6$ times.}  
  \label{ps}
\end{figure*}

Figure~\ref{ps} shows numerical estimates of the phase-space regions obtained 
for each regime $S_M$ in the chain of coupled maps defined in Sec.~\ref{mod}.  
The regime $S_0$ (or the ordered regime) is associated to region
localized close to the border of the KAM island of the uncoupled case
(compare to Fig.~\ref{ps}a). Points which belong to the regime $S_1$  
are closer to the center of the torus from the
uncoupled case. This suggests that when trajectories are inside the 
region related to regime $S_1$, they more likely 
penetrate inside the torus from the uncoupled case. In the chaotic sea both 
regimes $S_1$ and $S_2$ are visible. These results are
naturally understood in the perturbative limit (small coupling $\xi\ll
1$). {The regime} $S_0$ corresponds to $\lambda_i^{(100)} \approx 0$ for every $i=1,\ldots,N$, 
which is expected when the trajectory is stuck close to the $N$-dimensional 
tori built as the product of the $1$-dimensional tori of the uncoupled 
maps. In contrast, $S_M$ for $M>0$ implies that at least one
{FTLE $\lambda_i^{(\omega)} \gg 0$} and therefore the trajectory
projected in one map can be both in the chaotic and regular regions 
({\it e.g.}, $S_1$ for $N=2$ can be obtained {from 
$\lambda_1^{(\omega)} \gg 0,~\lambda_2^{(\omega)}
  \approx 0$ or from $\lambda_1^{(\omega)} \approx
  0,~\lambda_2^{(\omega)} \gg 0$)}. Altogether, these observations confirm that our 
method allows for a 
meaningful division of the chaotic component of the phase-space and can thus be 
used to identify regions of interesting dynamics. {In the case partial  barriers exist
  inside the chaotic component  -- such as in area-preserving
  maps with mixed phase space~\cite{Meiss} -- we expect the regions obtained through our
  method to depend weakly on  $\omega$ and to coincide with those obtained from the
  partial barriers.} 

\section{Results}
\label{nr}
In this Section we apply the Lyapunov time-series methodology described in 
Sec.~\ref{methods} to the $2N$-dimensional system defined in Sec.~\ref{mod}. 
We compute and interpret four basic properties of the method: the total time 
spent in each regime (residence time), the transition between regimes, the 
consecutive time in each regime, and the scaling of Lyapunov exponents. 

\subsection{Residence time in each regime}
\label{stpt}
The first and most basic quantity we measure is the probability $P(S_M)$ of 
finding the trajectory in each regime, defined as the fraction of the total 
time $t_L$ that {$\boldsymbol{x}_t \in S_M$} ({\it i.e.} $P(S_M) = 
\sum_{t=0}^{t_L} \delta_{t\in S_M}/t_L$, where $\delta_{t\in S_M}=1$ if 
$t\in S_M$ and $\delta_{t\in S_M}=0$ otherwise). 
\begin{figure}[!ht]
  \centering
  \includegraphics*[width=0.95\columnwidth]{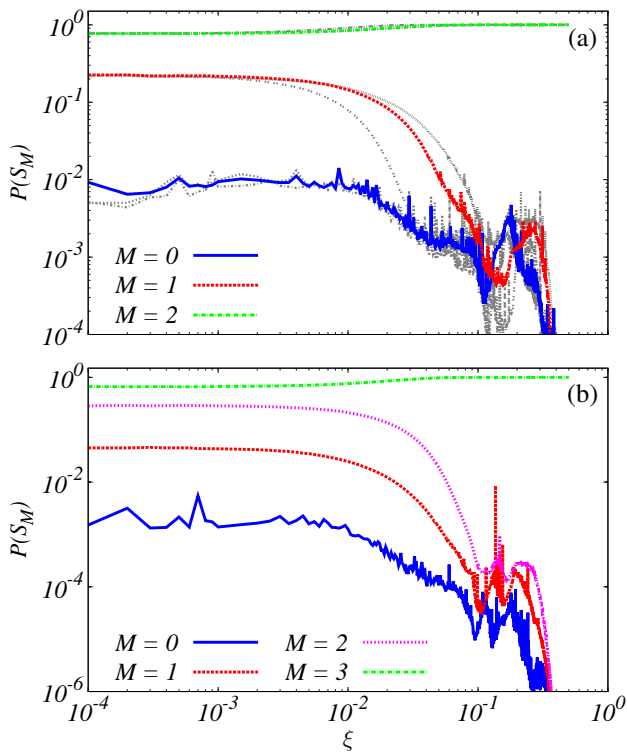}
  \caption{(Color online) Residence time in each regime $S_M$. (a) 
  $N=2$ with $\varepsilon_1=0.1$ and $\varepsilon_2=0.05$. (b) $N=3$
  with $\varepsilon_1=0.1$, $\varepsilon_2=0.08,$ and $\varepsilon_3=0.06$. 
 {In (a) the values obtained with $\omega=100$ are compared (gray curves) 
with results for $\omega=50$ and  $\omega=500$. Only for the case $M=1$ the 
gray curves (right for $\omega=50$ and left for $\omega=500$) show a shift 
in the $x$-axis $\xi$.}  Estimations for each $\xi$ are based on a
 trajectory with length  $t_L=10^{10}$.}  
  \label{commot}
\end{figure}

Figure \ref{commot} shows the probabilities $P(S_M)$ for the map with
$N=2,3$ as a function of the coupling strength $\xi$. We now explain
the behavior of $P(S_M)$ with $\xi$ by discussing the effect of
coupling $\xi$ on the phase-space regions associated to $S_M$, as
defined in Sec.~\ref{ph}. By the  ergodicity of
$\Gamma_{\text{chaos}}$, $P(S_M)$ corresponds to the (normalized)
volume of the region related to regime $S_M$ in the phase-space      
\begin{equation}
  \label{eq.Smu}
  P(S_M) = \frac{\mu(S_M)}{\mu(\Gamma_{\text{chaos}})}
  =\frac{\mu(S_M)}{1-\mu(\Gamma_{\text{tori}})}. 
\end{equation}
The results of Fig.~\ref{commot} show that the chaotic region is the  
largest region in phase-space for any coupling, while the region
associated to $S_{1}$ has a larger volume than $S_{0}$ for couplings
$\xi\lesssim 1.3\times 10^{-1}$. For larger $\xi$ we see oscillations
with a local maximum close to $\xi\sim 2\times10^{-1}$ for the cases
$M=0$ and $M=1$.  

We now interpret the $\xi$ dependence observed in Fig.~\ref{commot} by
arguing how the different terms in Eq.~(\ref{eq.Smu}) vary with
$\xi$. We denote by $\mu(U_j)$ the measure of tori for the $j$-th map
with control parameter $K_j$ in the uncoupled case $\xi=0$ (which we
assume to be approximately equal to the measure of the KAM
islands). For small coupling $\xi \approx 0$ we expect that most tori
of the uncoupled maps to survive and therefore: 
\begin{itemize}
  \item $\mu(\Gamma_{\text{tori}})\approx\prod_{j=1}^{N} \mu(U_j)$,
    which in the simple case of $\mu(U_j)=\mu$ for all $j$ reduces to
    $\mu(\Gamma_{\text{tori}})\approx \mu^N$.

  \item {$\mu(S_{M=0})$} corresponds to a small volume around
    $\Gamma_{\text{tori}}$, \textit{i.e.}
    $\mu(S_{M=0})\sim\mu(\Gamma_{\text{tori}})\approx
    P(S_0)/(1+P(S_0))$. 

  \item  {For $\mu(S_{M\ne0})$ we have that} 
     $N-M$ maps are in their corresponding KAM island
    (with probability $\mu(U_j)$) and $M$ maps in the chaotic area
    (with probability $1-\mu(U_k)$). {For example, for $N=3$ and $M=2$
    we have that 
\begin{eqnarray}\mu(S_{M=2}) &=& \mu(U_1)[1-\mu(U_2)][1-\mu(U_3)]\cr
    &+&\mu(U_2)[1-\mu(U_1)][1-\mu(U_3)]\cr
&+&\mu(U_3)[1-\mu(U_1)][1-\mu(U_2)].
\nonumber
\end{eqnarray}
    In general this} leads to 
    $$\mu(S_M) \approx \sum_{j_1} \ldots \sum_{j_M}
    \prod_{j\in\{j_1,\ldots,j_M\}} (1-\mu_j)
    \prod_{j\notin\{j_1,\ldots, j_M\}} \mu_j,$$ where the last product
    is over all $j=1, \ldots, N$ except $j\in\{j_1,\ldots,j_M\}$. In
    the simple case of $\mu(U_j)=\mu$, it reduces to 
    $\mu(S_M) \approx \binom{N}{M} \mu^{N-M} (1-\mu)^M.$  
\end{itemize}

We now consider the effect of growing $\xi$. In the spirit of the KAM theorem, 
the tori of the coupled maps (generated as the product of the $N$ maps) are 
expected to be robust to small couplings $\xi$, which act as a perturbation. 
This explains why the curves in Fig.~\ref{commot} are essentially flat for 
small $\xi$. Increasing $\xi$ even further, the nonlinearity of the system 
increases and therefore $\mu(\Gamma_{\text{tori}})$ is expected to decrease 
($\mu(\Gamma_{\text{tori}})\rightarrow 0$ for $\xi \gg 0$). This reduction of 
the tori leads to an increase in the denominator of Eq.~(\ref{eq.Smu}) and 
explains the observed tendency of reduction of $P(S_M)$ for all regions related 
to stickiness ($M<N$). Indeed, for $\xi>0.5$ no signature of tori or stickiness 
was detected numerically and $P(S_{M=N})=1$. The nontrivial dependencies of 
$P(S_{M<N})$ in Fig.~\ref{commot} appear at {$\xi\sim 2\times10^{-1}$} values, 
close to the values of $\xi$ for which the last tori disappear (see also Fig. 
$7.2$ in Ref.~\cite{AltmannThesis}). In this regime the volume of the tori is 
already negligible $\mu(\Gamma_{\text{tori}})\gtrsim 0$ but stickiness is still 
effective (notice that even zero measure non-hyperbolic sets can lead to 
stickiness \cite{amk06,Manchein}). The denominator in Eq.~(\ref{eq.Smu}) is 
therefore {$1-\mu(\Gamma_{\text{tori}})\sim 1$}, not significantly affected by 
further increases of $\xi$, and therefore not driving the reduction of $P(S_{M<N})$. 
Small variation of a control parameter of the system (in this case $\xi$) are 
known to lead to sensitive creation and destruction of tori, with non-trivial
dependencies on the stickiness~\cite{ZaslavskyPhysicsReports}. We can thus 
expect that -- close to the disappearance of the tori -- the small volume of 
stickiness regions $\mu(S_{M<N})$ to fluctuate with $\xi$ leading even to an 
increase with $\xi$. It is interesting to note that this non-trivial increase 
with $\xi$ appears for $P(S_{M=0})$ in Fig.~\ref{commot} precisely when the curves
$P(S_{0<M<N})$ show a sharp decreasing fluctuation. This suggests an exchange 
between measure of different sticky regions associated to regimes
$S_{M<N}$, without interference of the much larger fully chaotic
component $S_{M=N}$.   

\subsection{Transitions between regimes}

\begin{figure}[!ht]
  \centering
 \includegraphics*[width=0.95\columnwidth]{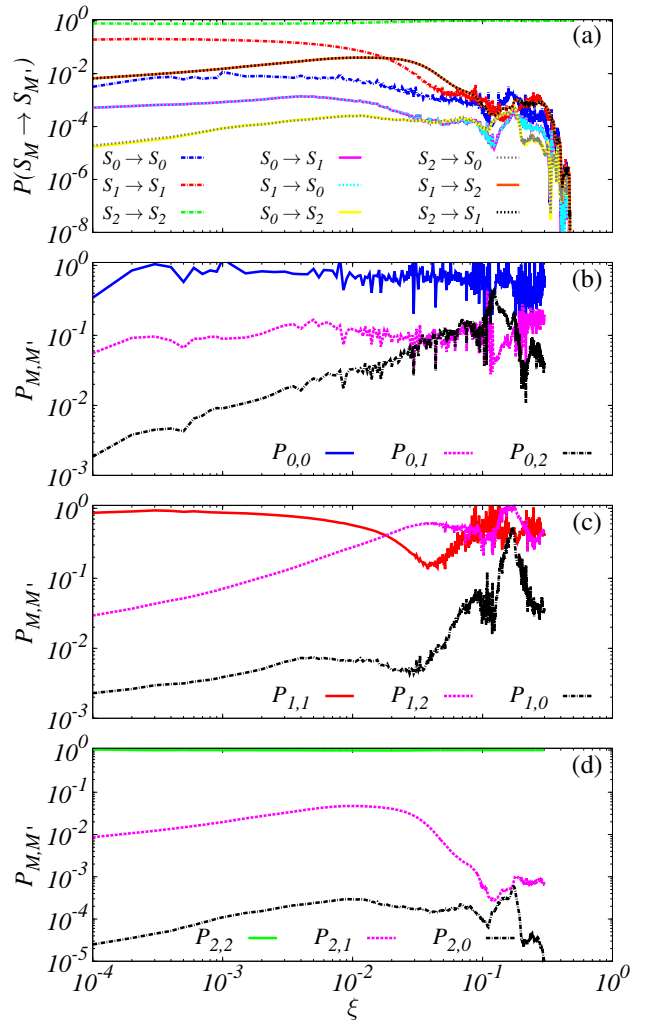}
  \caption{(Color online) Transition between regimes $S_M$ as a function of the 
  coupling strength $\xi$. (a) Transition probability $P(S_M \to S_{M'})$; (b-d) 
  Conditional probability $P_{M,M'}$ defined in Eq.~(\ref{nullmodel}) of moving 
  to $M'$ given that the trajectory was at $M$. Estimations for each $\xi$
  based on a single trajectory with length $t_L=10^{10}$ in the case of $N=2$ 
  coupled maps and $\varepsilon_1=0.1$ and $\varepsilon_2=0.05$.} 
  \label{prob1}
\end{figure}
We now focus on the transition between regimes. The simplest analysis correspond 
to the two-time (joint) probability $P(S_M \to S_{M'})$, computed as the fraction 
of the total trajectory time $t_L$ that $\boldsymbol{x}_t\in S_M$ and 
$\boldsymbol{x}_{t+1} \in S_{M'}$. The probabilities considered in the previous 
section can be obtained as $\sum_{S_M} P(S_M \to S_{M'})=P(S_{M'})$ and 
$\sum_{S_{M'}} P(S_M \to S_{M'})=P(S_M)$. Figure~\ref{prob1}(a) shows the 
dependence of $P(S_M \to S_{M'})$ on $\xi$ for our model. We notice that $P(S_M
\to S_{M'})$ is equal to $P(S_{M'} \to S_{M})$. This is expected considering 
that the system is ergodic, volume preserving, and time-reversible. The 
dependence of $P(S_M \to S_{M'})$ on $\xi$ follows a similar pattern observed 
for $P(S_M)$ in Fig.~\ref{commot}. More information is obtained from the 
conditional probability 
\begin{equation}
  \label{nullmodel}
  P_{M,M'} \equiv P(S_M \rightarrow
  S_{M^{\prime}}|S_M)\equiv\dfrac{P(S_M \rightarrow
    S_{M^{\prime}})}{P(S_M)}, 
\end{equation}
which quantifies the probability that trajectories at $S_M$ will move to 
$S_{M^{\prime}}$. The results shown in Fig.~\ref{prob1}(b-d) show for all 
$S_M$ that (i) persistence in the same $S_M$ ($P(S_M\leftrightarrow S_{M}|S_M)$) 
is dominant and (ii) the most likely transitions occur between neighboring 
regimes ({\it e.g.}, $P_{2,1}>P_{2,0}$). The only (slight) deviations of this 
picture happen for large values of $\xi$, close to the disappearance of the
KAM island. Altogether, these results confirm that in the perturbative
regime ($\xi \ll 1$) stickiness happens approaching the region of
regular motion of different maps one after the other (in opposite to a
direct approach from $S_0$ to $S_{M=N}$).

\subsection{Consecutive time in each regime} 
\label{rt}
The results of the previous section confirm that residence in the same regime 
is the dominant behavior. This motivates us to study the time $\tau_M$ spent 
consecutively in a regime $S_M$ ({\it i.e.}, $\tau_M$ is the time between two 
consecutive transitions between different regimes, the first to $S_M$ and the 
second out of $S_M$). In a trajectory of length $t_L$ we collect a series of 
$\tau_M$. We are mainly interested in the probability distribution $P(\tau_M)$ 
(or, equivalently its cumulative $P_{cum} \equiv \sum_{\tau'_M=\tau_M}^{\infty} 
P(\tau'_M)$) for different $S_M$'s in the limit $t_L \rightarrow \infty$. These 
distributions should be compared to the distribution of recurrence times $\tau$, 
defined as the time between two successive entries to a pre-defined recurrence 
region (usually taken in the fully chaotic component of the phase-space). Events 
in the tails of $P(\tau)$ are associated to times for which the trajectory is 
stuck to the non-hyperbolic components of the phase-space and $P(\tau)$ is the 
traditional method to quantify stickiness in Hamiltonian systems~\cite{Chir-Shep,
Cristadoro,Artuso,AltmannKantz}.
\begin{figure}[!ht]
  \centering
  \includegraphics*[width=1\columnwidth]{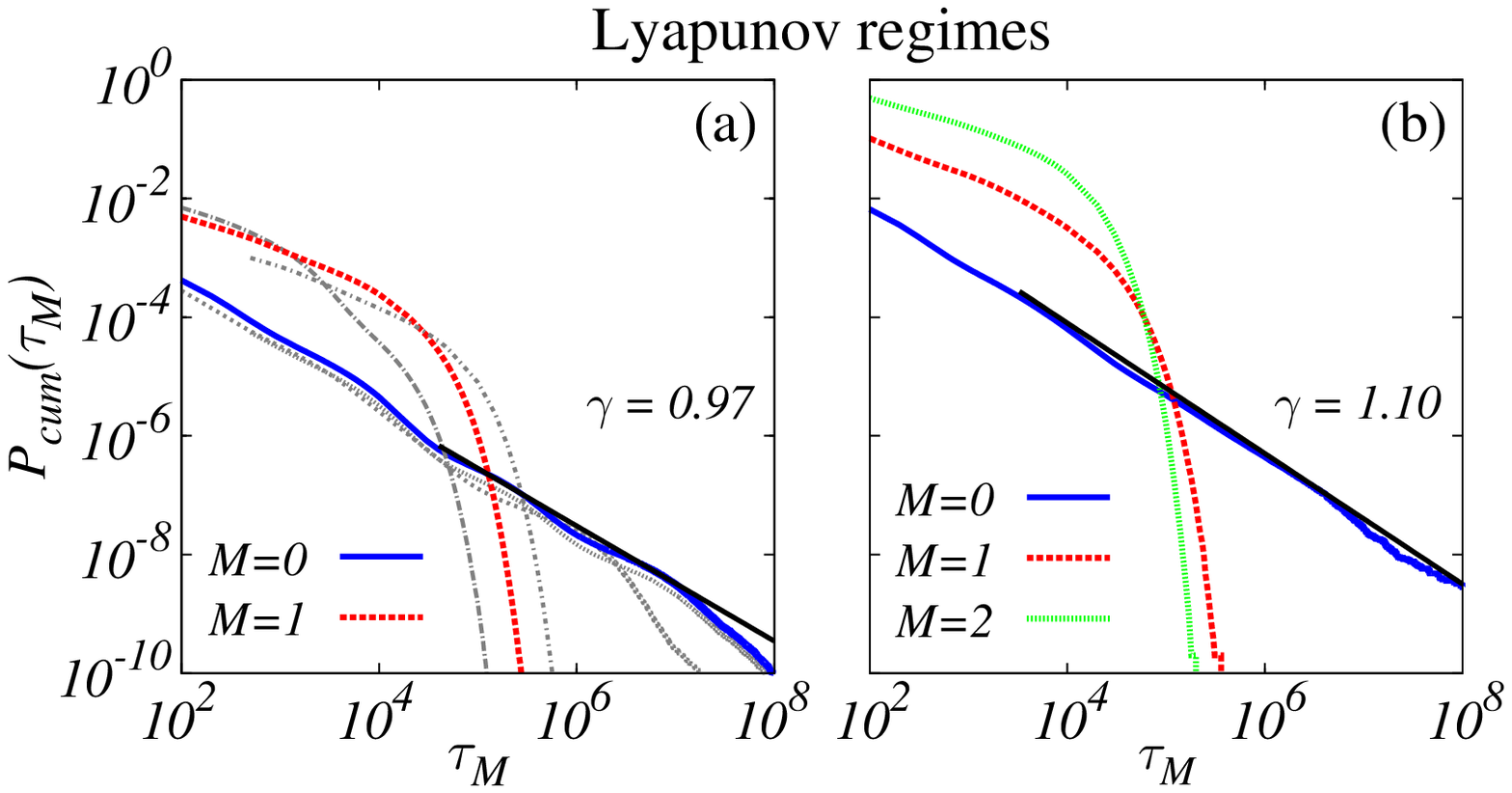}
  \includegraphics*[width=1\columnwidth]{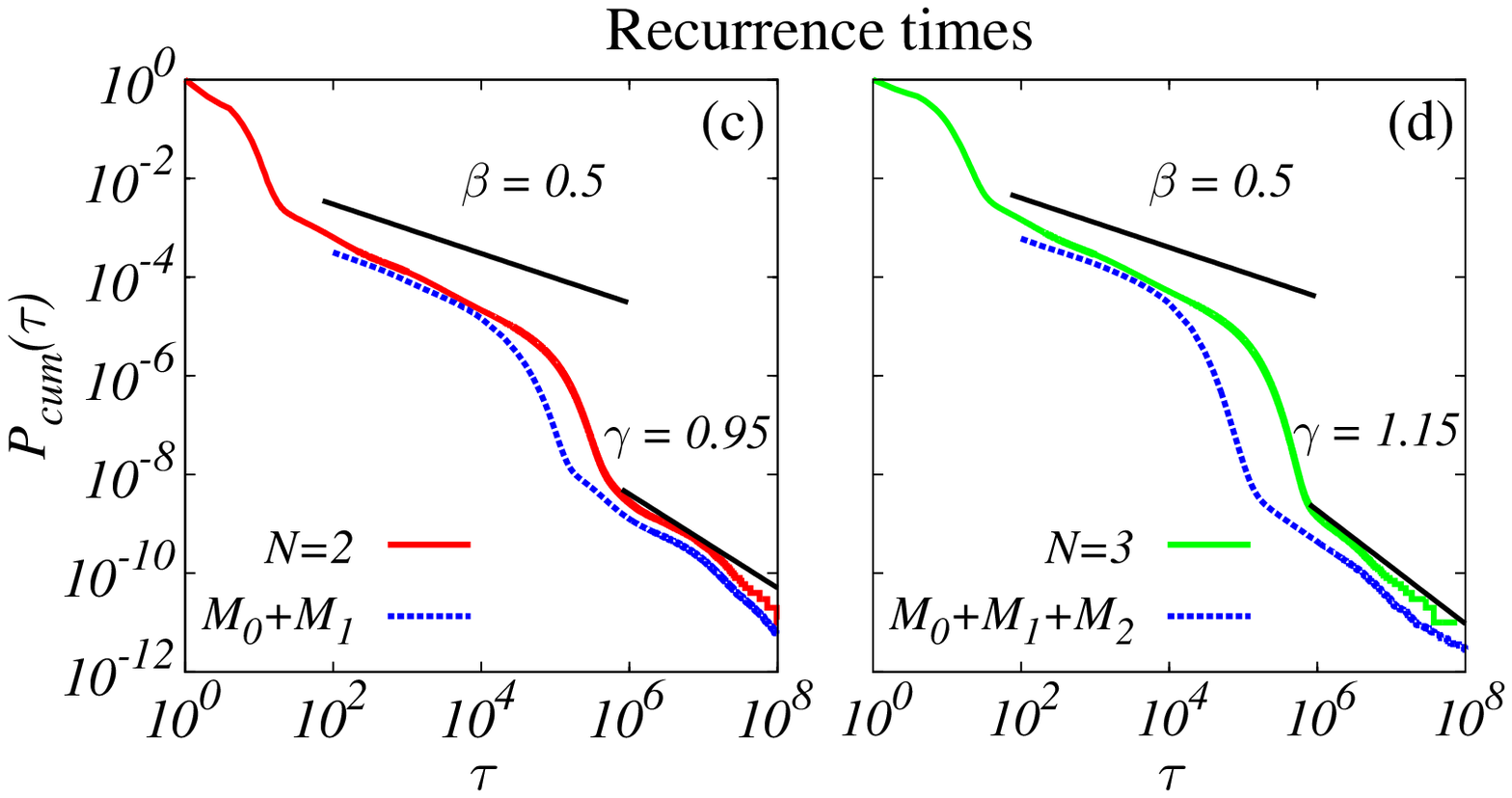}
  \caption{(Color online) Comparison between our method and the analysis based 
  on recurrence time. The cumulative distribution $P_{{cum}}(\tau_M)$  of times 
  $\tau_M$ is shown for each regime $S_{M<N}$ for $\omega=100$ and (a) $N=2$ 
  and (b) $N=3$.  {In (a) the gray curves show  results for 
  $\omega=50$ and $\omega=500$.  Only for the case $M=1$ the gray curves 
(left for $\omega=50$ and right for $\omega=500$) show a shift in the 
$x$-axis $\xi$.
} The 
  cumulative distribution $P_{{cum}}(\tau)$ for recurrence times $\tau$ to a 
  region in the chaotic component of the phase-space (in
  $S_{M=N}$) for (c) $N=2$ and (d) $N=3$. For comparison, in
  panels (c) we show the  results obtained combining the
  {normalized} curves for $M_0+M_1$ {(blue dotted line: divided by
    $1.7\times 10^3$ for convenience of scale)} of panel (a), and in (d) the
  {normalized} curves for $M_0+M_1+M_2$ {(blue dotted line: divided by
    $10^3$)} of 
  panel (b). Results obtained using maps of Sec.~\ref{mod} with  
  $\xi = 10^{-3}$, $\varepsilon_1=0.1$ and $\varepsilon_2=0.05$ for the
  case $N=2$ and $\varepsilon_1=0.1$, $\varepsilon_2=0.08$ and
  $\varepsilon_3=0.06$ for the case $N=3$.}
\label{pcum}
\end{figure}

The numerical simulations in Fig.~\ref{pcum} confirm that the distribution 
obtained summing $P_{cum}(\tau_M)$ for ordered and semi-ordered regimes 
(or $S_{M<N}$) is equivalent to cumulative distribution $P_{cum}(\tau)$ 
obtained using recurrences. This is in agreement with the association of 
long consecutive times in regimes of ordered and semi-ordered motion to long 
recurrence times.  Looking at the individual distributions 
$P_{cum}(\tau_M)$ provide valuable additional information on the sticky motion. 
For semi-ordered motion {(when $0<M<N$)} we observe an exponential
tail after an intermediate decay with scaling $\beta \approx
0.5$. This behavior confirms the interpretation given in
Ref.~\cite{AltmannKantz}. More interestingly, the $M=0$ case shows an
asymptotic algebraic decay which characterize stickiness. While the
scaling is compatible with the results obtained using recurrence time,
$P_{cum}(\tau_{M=0})$ obtained in our methodology provides a better
characterization of the scaling (over several orders of magnitudes)
and allows for an independent analysis of the different regimes. These
properties are essential when dealing with high-dimensional systems
(which may contain different pre-asymptotic regimes) and for an
accurate estimation of the stickiness exponent $\gamma$.  {Finally, panel (a) in
  Fig.~\ref{pcum} shows that all  decays discussed above remain (qualitatively) the same
  for different choices of  $\omega$, with the curve for  $M=1$ showing the largest
  sensitivity on $\omega$ (as in Fig.~\ref{commot}(a). }

\subsection{Scaling of Lyapunov exponents}
\label{sle}

So far we have focused at the temporal properties of the time series of 
FTLEs~{$\lambda_i^{(\omega)}$} and how they change with the coupling 
strength~$\xi$. We now consider how the value of the Lyapunovs respond to an
external perturbation, which in our case is the coupling to the other maps. It 
is known that the largest exponent 
{$\lambda_1^{(\infty)}$} is extremely sensitive to perturbation. More specifically, 
Daido's relation~\cite{Daido1,ZillmAhlersPik} states that for small couplings 
$\xi$ to another chaotic system, a universal logarithmic singularity is observed,
\begin{equation}
  \lambda_i^{(\omega\rightarrow\infty)} -
  \Lambda_i^{(\omega\rightarrow\infty)} \approx \dfrac{c}{|\ln(\xi)|}, 
  \label{daido}
\end{equation}
where $\Lambda_i^{(\omega\rightarrow\infty)}$ are the unperturbed
Lyapunov exponents and $c$ is a constant and $i=1,\ldots,N$. 
This  relation is valid for {totally 
chaotic systems
  and} a small mismatch  between Lyapunov exponents of the uncoupled
  systems {compared to their fluctuations}~\cite{ZillmAhlersPik}.
Here we investigate the {relation  $\lambda_i^{(\omega)} -
\Lambda_i^{(\omega)}$ as a function of $\xi$, for distinct values of $\omega$}
and 
different regimes $S_M$.  {To this end we compute the temporal averages of 
the FTLEs $\langle \lambda^{(\omega)}_i \rangle$ for 
times $t$ such that $\{\lambda_i^{(\omega)}\}\in S_M$.}

\begin{figure}[!h]
  \centering
  \includegraphics*[width=1\columnwidth]{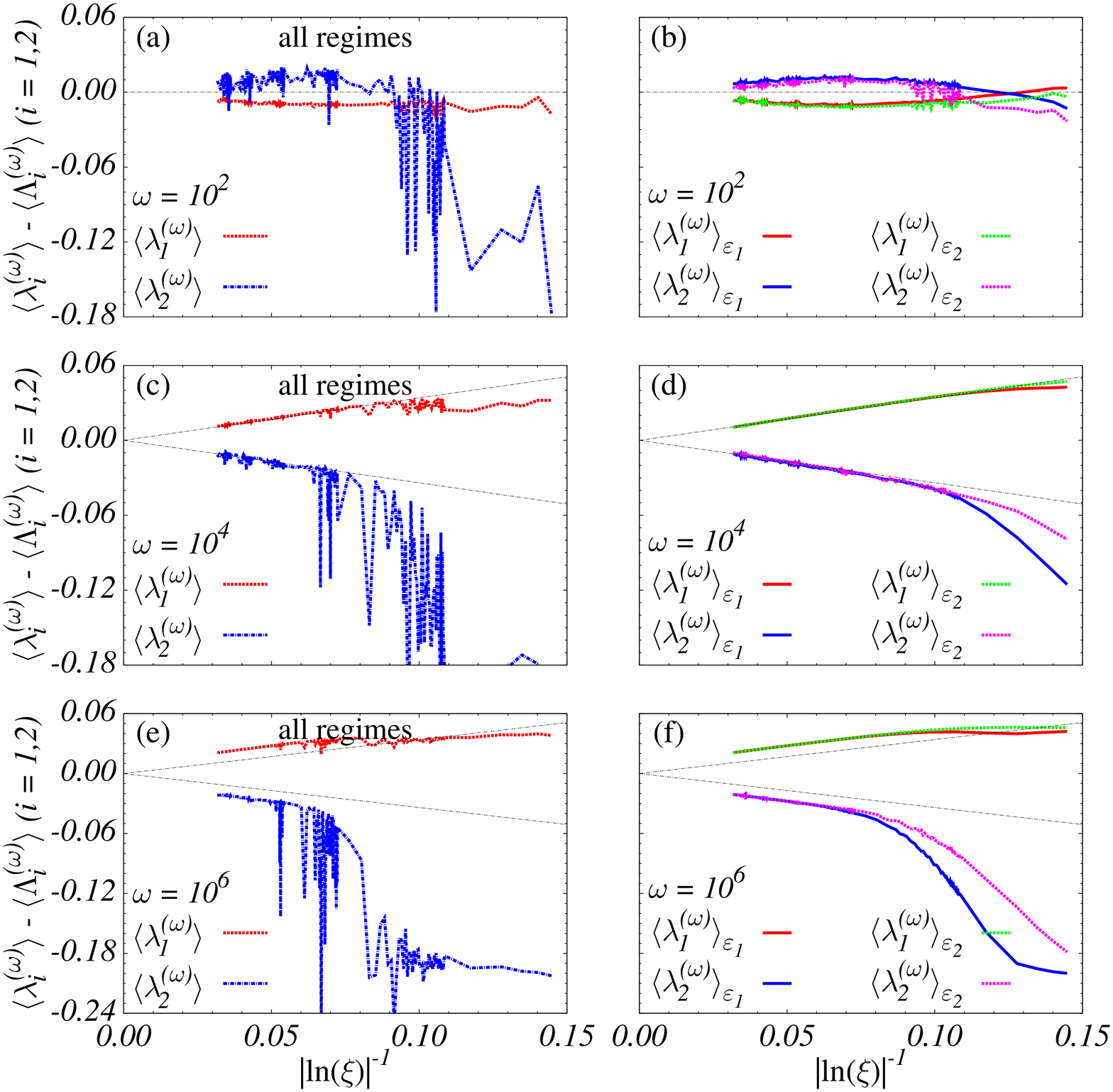}
  \caption{(Color online) Sensitivity of the FTLEs for small
    couplings. The difference between the finite-time Lyapunov exponent 
    in the coupled ({$\langle\lambda_i^{(\omega)}\rangle$}) and
    uncoupled ({$\langle\Lambda_i^{(\omega)}\rangle$}) maps as a
    function of {$|\ln(\xi)|^{-1}$}, where $\xi$ is the coupling
    strength. Results are shown for $N=2$ {($i=1,2$)} and different 
    time windows $\omega$ (a)-(b) $10^2$, {(c)-(d) $10^4$}, and 
    {(e)-(f) $10^6$}. Black dashed lines in (c)-(f) are the expected linear
    behaviour, consistent with Eq.~(\ref{daido}). Panels in the left
    column (ace) were computed for the full time series, while on the
    right column (bdf) only {FTLEs} in the regime $S_2$ were
    used. The different colors correspond to different choices of
    threshold imposed to define the {FTLE}:  
    {$\langle\lambda^{(\omega)}_i\rangle_{\varepsilon_1}$}
    uses {$\varepsilon_1=0.1\langle\lambda_i^{(\omega)}\rangle$}
    while {$\langle\lambda^{(\omega)}_i\rangle_{\varepsilon_2}$} uses 
    {$\varepsilon_2=0.9\langle\lambda_i^{(\omega)}\rangle$},
    where {$\langle\lambda_i^{(\omega)}\rangle$}  
    is computed over the full time series ({left column}). } 
  \label{scal}
\end{figure}

Our numerical simulations reported in Fig.~\ref{scal} show that 
small values of $\omega$ lead to a
situation in which 
{$\langle\lambda_i^{(\omega)}\rangle\approx\langle\Lambda_i^{(\omega)}\rangle$}
at a finite value of $\xi$ (Figs.~\ref{scal}(b) and (d)), 
{while larger values of $\omega$ lead to situations in which
{$\langle\lambda_i^{(\omega)}\rangle\neq\langle\Lambda_i^{(\omega)}\rangle$}
for any $\xi$. These results depend crucially on our choice to impose the order of
$\lambda_i^{(\omega)}$ for all $t$, as discussed in Sec.~\ref{regions}. This makes the average over
the trajectory time $\langle \lambda_i^{(\omega} \rangle$ to be 
$\omega$-dependent and different from the average over the Lyapunov time $\lambda_i^{(\omega\rightarrow\infty)}$.}  Applying the analysis
without the division in regimes $S_M$ leads to strongly fluctuating results
(Figs.~\ref{scal}(a,c,e)). {Much smoother results (Figs.~\ref{scal}(b,d,f)) are
  obtained when we apply our method and compute $\langle \lambda_i^{(\omega)} \rangle$ only
  for $t$ in the fully chaotic regime $S_N$.} 
{Looking at these smoother results we observe that the difference in Lyapunovs
  scales as $1/|\ln{\xi}|$,  but that even for $\omega \rightarrow \infty$  the sticky
  motion leads to a deviation  from Daido's relation~(\ref{daido}) (curves are shifted vertically).}

\section{Conclusions}
\label{cc}

{In summary, we have proposed a method to characterize the
dynamics of Hamiltonian systems with mixed phase-space based on time
series of finite-time Lyapunov exponents. Using this method it is 
possible to define and study with high accuracy  the time evolution of
regimes of ordered, semi-ordered, and totally chaotic motion. This allows 
for an individualized characterization of the different stickiness
mechanisms, improving alternative methods based on the statistics of
recurrence times or on the distribution of finite-time Lyapunov
exponents.} 

{We applied our method to a chain of coupled standard maps and
  showed how the frequency of different regimes  -- and the transition
  probabilities between them -- are related to the volume of different
  phase-space regions.  Using the consecutive time in distinct regimes
  we have reproduced previous results obtained using recurrence times
  and showed that our method allows for a significant improvement in
  the characterization of the sticky motion ({\it e.g.}, in the
  determination of the scaling exponents). This indicates that 
our method can be used to characterize stickiness in general
high-dimensional systems and is particularly suited for cases in which
different regions of sticky motion coexist.} We have also shown that the 
dependence on the coupling strength of the largest Lyapunov exponents,
after conveniently using our procedure, 
{tend to follow only the qualitative} 
universal properties of fully chaotic system.  

Results obtained in a simple chain of standard maps confirm that our 
methodology can be applied to high-dimensional systems and problems of current 
interest, such as controlling Fermi acceleration \cite{ibere12}, galactic
models \cite{contopoulos10-1}, and plasma physics \cite{negrete05}.  Another 
example of application is to associate each regime $S_M$ with effective 
Hamiltonian functions, a procedure used to reproduce the complicated
dynamics of kicking  
electrons \cite{rost12-1} or the high harmonic generation in laser-assisted 
collisions \cite{rost12-2}.

\section*{Acknowledgments}

CM and RMS thank CNPq, CAPES and FAPESC and MWB thanks CNPq for
financial support and MPIPKS in the framework of the Advanced Study Group
on Optical Rare Events. CM also thanks Eduardo G. Altmann for the
financial support and hospitality at the MPIPKS. EGA thanks D.~Paz\'o
for suggesting the analysis performed in Sec.~\ref{sle}.



\begin{thebibliography}{37}
\expandafter\ifx\csname natexlab\endcsname\relax\def\natexlab#1{#1}\fi
\expandafter\ifx\csname bibnamefont\endcsname\relax
  \def\bibnamefont#1{#1}\fi
\expandafter\ifx\csname bibfnamefont\endcsname\relax
  \def\bibfnamefont#1{#1}\fi
\expandafter\ifx\csname citenamefont\endcsname\relax
  \def\citenamefont#1{#1}\fi
\expandafter\ifx\csname url\endcsname\relax
  \def\url#1{\texttt{#1}}\fi
\expandafter\ifx\csname urlprefix\endcsname\relax\def\urlprefix{URL }\fi
\providecommand{\bibinfo}[2]{#2}
\providecommand{\eprint}[2][]{\url{#2}}

\bibitem[{\citenamefont{Lichtenberg and Lieberman}(1992)}]{Lichtenberg}
\bibinfo{author}{\bibfnamefont{A.~J.} \bibnamefont{Lichtenberg}}
  \bibnamefont{and} \bibinfo{author}{\bibfnamefont{M.~A.}
  \bibnamefont{Lieberman}}, \emph{\bibinfo{title}{Regular and Chaotic
  Dynamics}} (\bibinfo{publisher}{Springer-Verlag}, \bibinfo{address}{New
  York}, \bibinfo{year}{1992}).

\bibitem[{\citenamefont{Meiss}(1992)}]{Meiss}
\bibinfo{author}{\bibfnamefont{J.~D.} \bibnamefont{Meiss}},
  \bibinfo{journal}{Rev. Mod. Phys.} \textbf{\bibinfo{volume}{64}},
  \bibinfo{pages}{795} (\bibinfo{year}{1992}).

\bibitem[{\citenamefont{Chirikov and Shepelyansky}(1984)}]{Chir-Shep}
\bibinfo{author}{\bibfnamefont{B.~V.} \bibnamefont{Chirikov}} \bibnamefont{and}
  \bibinfo{author}{\bibfnamefont{D.~L.} \bibnamefont{Shepelyansky}},
  \bibinfo{journal}{Physica D} \textbf{\bibinfo{volume}{13D}},
  \bibinfo{pages}{395} (\bibinfo{year}{1984}).

\bibitem[{\citenamefont{Artuso}(1999)}]{Artuso}
\bibinfo{author}{\bibfnamefont{R.}~\bibnamefont{Artuso}},
  \bibinfo{journal}{Physica D} \textbf{\bibinfo{volume}{131}},
  \bibinfo{pages}{68} (\bibinfo{year}{1999}).

\bibitem[{\citenamefont{Zaslavsky}(2002)}]{ZaslavskyPhysicsReports}
\bibinfo{author}{\bibfnamefont{G.~M.} \bibnamefont{Zaslavsky}},
  \bibinfo{journal}{Physics Reports} \textbf{\bibinfo{volume}{371}},
  \bibinfo{pages}{461} (\bibinfo{year}{2002}).

\bibitem[{\citenamefont{Altmann}(2007)}]{AltmannThesis}
\bibinfo{author}{\bibfnamefont{E.~G.} \bibnamefont{Altmann}}, Ph.D. thesis,
  \bibinfo{school}{Max Planck Institut f\"{u}r Physik Komplexer Systeme}
  (\bibinfo{year}{2007}).

\bibitem[{\citenamefont{Artuso and Manchein}(2009)}]{Manchein}
\bibinfo{author}{\bibfnamefont{R.}~\bibnamefont{Artuso}} \bibnamefont{and}
  \bibinfo{author}{\bibfnamefont{C.}~\bibnamefont{Manchein}},
  \bibinfo{journal}{Phys. Rev. E} \textbf{\bibinfo{volume}{80}},
  \bibinfo{pages}{036210} (\bibinfo{year}{2009}).

\bibitem[{\citenamefont{Cristadoro and Ketzmerick}(2008)}]{Cristadoro}
\bibinfo{author}{\bibfnamefont{G.}~\bibnamefont{Cristadoro}} \bibnamefont{and}
  \bibinfo{author}{\bibfnamefont{R.}~\bibnamefont{Ketzmerick}},
  \bibinfo{journal}{Phys. Rev. Lett.} \textbf{\bibinfo{volume}{100}},
  \bibinfo{pages}{184101} (\bibinfo{year}{2008}).

\bibitem[{\citenamefont{Altmann and Kantz}(2007)}]{AltmannKantz}
\bibinfo{author}{\bibfnamefont{E.~G.} \bibnamefont{Altmann}} \bibnamefont{and}
  \bibinfo{author}{\bibfnamefont{H.}~\bibnamefont{Kantz}},
  \bibinfo{journal}{Europhys. Lett.} \textbf{\bibinfo{volume}{78}},
  \bibinfo{pages}{10008} (\bibinfo{year}{2007}).

\bibitem[{\citenamefont{Kantz and Grassberger}(1987)}]{GrassbergerKantz}
\bibinfo{author}{\bibfnamefont{H.}~\bibnamefont{Kantz}} \bibnamefont{and}
  \bibinfo{author}{\bibfnamefont{P.}~\bibnamefont{Grassberger}},
  \bibinfo{journal}{Phys. Lett. A} \textbf{\bibinfo{volume}{123}},
  \bibinfo{pages}{437} (\bibinfo{year}{1987}).

\bibitem[{\citenamefont{Szezech et~al.}(2005)\citenamefont{Szezech, Lopes, and
  Viana}}]{Viana}
\bibinfo{author}{\bibfnamefont{J.~D.} \bibnamefont{Szezech}},
  \bibinfo{author}{\bibfnamefont{S.~R.} \bibnamefont{Lopes}}, \bibnamefont{and}
  \bibinfo{author}{\bibfnamefont{R.~L.} \bibnamefont{Viana}},
  \bibinfo{journal}{Phys. Lett. A} \textbf{\bibinfo{volume}{335}},
  \bibinfo{pages}{394} (\bibinfo{year}{2005}).

\bibitem[{\citenamefont{Harle and Feudel}(2007)}]{Harle}
\bibinfo{author}{\bibfnamefont{M.}~\bibnamefont{Harle}} \bibnamefont{and}
  \bibinfo{author}{\bibfnamefont{U.}~\bibnamefont{Feudel}},
  \bibinfo{journal}{Chaos, Solitons $\&$ Fractals}
  \textbf{\bibinfo{volume}{31}}, \bibinfo{pages}{130} (\bibinfo{year}{2007}).

\bibitem[{\citenamefont{Laffargue et~al.}(2013)\citenamefont{Laffargue, Lam,
  Kurchan, and Tailleur}}]{New}
\bibinfo{author}{\bibfnamefont{T.}~\bibnamefont{Laffargue}},
  \bibinfo{author}{\bibfnamefont{K.-D. N.~T.} \bibnamefont{Lam}},
  \bibinfo{author}{\bibfnamefont{J.}~\bibnamefont{Kurchan}}, \bibnamefont{and}
  \bibinfo{author}{\bibfnamefont{J.}~\bibnamefont{Tailleur}},
  \bibinfo{journal}{Journal of Physics A: Mathematical and Theoretical}
  \textbf{\bibinfo{volume}{46}}, \bibinfo{pages}{254002}
  (\bibinfo{year}{2013}).

\bibitem[{\citenamefont{Manchein et~al.}(2012)\citenamefont{Manchein, Beims,
  and Rost}}]{cesar12}
\bibinfo{author}{\bibfnamefont{C.}~\bibnamefont{Manchein}},
  \bibinfo{author}{\bibfnamefont{M.~W.} \bibnamefont{Beims}}, \bibnamefont{and}
  \bibinfo{author}{\bibfnamefont{J.~M.} \bibnamefont{Rost}},
  \bibinfo{journal}{Chaos} \textbf{\bibinfo{volume}{22}},
  \bibinfo{pages}{033137} (\bibinfo{year}{2012}).

\bibitem[{\citenamefont{Manchein et~al.}(2014)\citenamefont{Manchein, Beims,
  and Rost}}]{cesar14}
\bibinfo{author}{\bibfnamefont{C.}~\bibnamefont{Manchein}},
  \bibinfo{author}{\bibfnamefont{M.~W.} \bibnamefont{Beims}}, \bibnamefont{and}
  \bibinfo{author}{\bibfnamefont{J.~M.} \bibnamefont{Rost}},
  \bibinfo{journal}{Physica A} \textbf{\bibinfo{volume}{400}},
  \bibinfo{pages}{186} (\bibinfo{year}{2014}).

\bibitem[{\citenamefont{Gaspard and Dorfman}(1995)}]{gd95}
\bibinfo{author}{\bibfnamefont{P.}~\bibnamefont{Gaspard}} \bibnamefont{and}
  \bibinfo{author}{\bibfnamefont{J.}~\bibnamefont{Dorfman}},
  \bibinfo{journal}{Phys. Rev. E} \textbf{\bibinfo{volume}{52}},
  \bibinfo{pages}{3525} (\bibinfo{year}{1995}).

\bibitem[{\citenamefont{Altmann et~al.}(2006)\citenamefont{Altmann, Motter, and
  Kantz}}]{amk06}
\bibinfo{author}{\bibfnamefont{E.~G.} \bibnamefont{Altmann}},
  \bibinfo{author}{\bibfnamefont{A.~E.} \bibnamefont{Motter}},
  \bibnamefont{and} \bibinfo{author}{\bibfnamefont{H.}~\bibnamefont{Kantz}},
  \bibinfo{journal}{Phys. Rev. E} \textbf{\bibinfo{volume}{73}},
  \bibinfo{pages}{026207} (\bibinfo{year}{2006}).

\bibitem[{\citenamefont{Artuso and Prampolini}(1998)}]{Artuso2}
\bibinfo{author}{\bibfnamefont{R.}~\bibnamefont{Artuso}} \bibnamefont{and}
  \bibinfo{author}{\bibfnamefont{A.}~\bibnamefont{Prampolini}},
  \bibinfo{journal}{Phys. Lett. A} \textbf{\bibinfo{volume}{246}},
  \bibinfo{pages}{407} (\bibinfo{year}{1998}).

\bibitem[{\citenamefont{Sala et~al.}()\citenamefont{Sala, Manchein, and
  Artuso}}]{sma14}
\bibinfo{author}{\bibfnamefont{M.}~\bibnamefont{Sala}},
  \bibinfo{author}{\bibfnamefont{C.}~\bibnamefont{Manchein}}, \bibnamefont{and}
  \bibinfo{author}{\bibfnamefont{R.}~\bibnamefont{Artuso}},
  \bibinfo{note}{arXiv:1410.4806}.

\bibitem[{\citenamefont{Contopoulos et~al.}(1978)\citenamefont{Contopoulos,
  Galgani, and Giorgilli}}]{Contopoulos}
\bibinfo{author}{\bibfnamefont{G.}~\bibnamefont{Contopoulos}},
  \bibinfo{author}{\bibfnamefont{L.}~\bibnamefont{Galgani}}, \bibnamefont{and}
  \bibinfo{author}{\bibfnamefont{A.}~\bibnamefont{Giorgilli}},
  \bibinfo{journal}{Phys. Rev. A} \textbf{\bibinfo{volume}{18}},
  \bibinfo{pages}{1183} (\bibinfo{year}{1978}).

\bibitem[{\citenamefont{Malagoli et~al.}(1986)\citenamefont{Malagoli, Paladin,
  and Vulpiani}}]{Malagoli}
\bibinfo{author}{\bibfnamefont{A.}~\bibnamefont{Malagoli}},
  \bibinfo{author}{\bibfnamefont{G.}~\bibnamefont{Paladin}}, \bibnamefont{and}
  \bibinfo{author}{\bibfnamefont{A.}~\bibnamefont{Vulpiani}},
  \bibinfo{journal}{Phys. Rev. A} \textbf{\bibinfo{volume}{34}},
  \bibinfo{pages}{1550} (\bibinfo{year}{1986}).

\bibitem[{\citenamefont{Mingzhou et~al.}(1990)\citenamefont{Mingzhou, Bountis,
  and Ott}}]{bountis}
\bibinfo{author}{\bibfnamefont{D.}~\bibnamefont{Mingzhou}},
  \bibinfo{author}{\bibfnamefont{T.}~\bibnamefont{Bountis}}, \bibnamefont{and}
  \bibinfo{author}{\bibfnamefont{E.}~\bibnamefont{Ott}},
  \bibinfo{journal}{Phys.~Lett.~A} \textbf{\bibinfo{volume}{151}},
  \bibinfo{pages}{395} (\bibinfo{year}{1990}).

\bibitem[{\citenamefont{Lange et~al.}(2010)\citenamefont{Lange, Richter, Onken,
  B\"acker, and Ketzmerick}}]{lange10}
\bibinfo{author}{\bibfnamefont{S.}~\bibnamefont{Lange}},
  \bibinfo{author}{\bibfnamefont{M.}~\bibnamefont{Richter}},
  \bibinfo{author}{\bibfnamefont{F.}~\bibnamefont{Onken}},
  \bibinfo{author}{\bibfnamefont{A.}~\bibnamefont{B\"acker}}, \bibnamefont{and}
  \bibinfo{author}{\bibfnamefont{R.}~\bibnamefont{Ketzmerick}},
  \bibinfo{journal}{Chaos} \textbf{\bibinfo{volume}{24}},
  \bibinfo{pages}{024409} (\bibinfo{year}{2010}).

\bibitem[{\citenamefont{Froeschlé}(1971)}]{Froeschle1}
\bibinfo{author}{\bibfnamefont{C.}~\bibnamefont{Froeschlé}},
  \bibinfo{journal}{Astrophys. Space Sci.} \textbf{\bibinfo{volume}{14}},
  \bibinfo{pages}{110} (\bibinfo{year}{1971}).

\bibitem[{\citenamefont{Froeschlé}(1972)}]{Froeschle2}
\bibinfo{author}{\bibfnamefont{C.}~\bibnamefont{Froeschlé}},
  \bibinfo{journal}{Astron. $\&$ Astrophys.} \textbf{\bibinfo{volume}{16}},
  \bibinfo{pages}{172} (\bibinfo{year}{1972}).

\bibitem[{\citenamefont{Dellnitz and Junge}(1997)}]{dellnitz97}
\bibinfo{author}{\bibfnamefont{M.}~\bibnamefont{Dellnitz}} \bibnamefont{and}
  \bibinfo{author}{\bibfnamefont{O.}~\bibnamefont{Junge}},
  \bibinfo{journal}{Int. J. Bif. Chaos} \textbf{\bibinfo{volume}{7}},
  \bibinfo{pages}{2475} (\bibinfo{year}{1997}).

\bibitem[{\citenamefont{Froyland and Padberg}(2009)}]{froyland09}
\bibinfo{author}{\bibfnamefont{G.}~\bibnamefont{Froyland}} \bibnamefont{and}
  \bibinfo{author}{\bibfnamefont{K.}~\bibnamefont{Padberg}},
  \bibinfo{journal}{Physica D} \textbf{\bibinfo{volume}{238}},
  \bibinfo{pages}{1507} (\bibinfo{year}{2009}).

\bibitem[{\citenamefont{Daido}(1984)}]{Daido1}
\bibinfo{author}{\bibfnamefont{H.}~\bibnamefont{Daido}},
  \bibinfo{journal}{Prog. Theor. Phys.} \textbf{\bibinfo{volume}{72}},
  \bibinfo{pages}{853} (\bibinfo{year}{1984}).

\bibitem[{Note1()}]{Note1}
Note1, \bibinfo{note}{without loss of generality we focus on the $N$ largest
  Lyapunov exponents because due to the symplectic character of Hamiltonian
  systems the others $N$ exponents are simply $\lambda _{N+1}=-\lambda _N,
  \lambda _{N+2}= -\lambda _{N-1}, \protect \ldots \lambda _{2N}=-\lambda _1$.}

\bibitem[{\citenamefont{Benettin et~al.}(1980)\citenamefont{Benettin, Galgani,
  Giorgilli, and Strelcyn}}]{bggs80}
\bibinfo{author}{\bibfnamefont{G.}~\bibnamefont{Benettin}},
  \bibinfo{author}{\bibfnamefont{L.}~\bibnamefont{Galgani}},
  \bibinfo{author}{\bibfnamefont{A.}~\bibnamefont{Giorgilli}},
  \bibnamefont{and} \bibinfo{author}{\bibfnamefont{J.-M.}
  \bibnamefont{Strelcyn}}, \bibinfo{journal}{Meccanica}
  \textbf{\bibinfo{volume}{15}}, \bibinfo{pages}{09} (\bibinfo{year}{1980}).

\bibitem[{\citenamefont{Wolf et~al.}(1985)\citenamefont{Wolf, Swift, Swinney,
  and Vastano}}]{wolf85}
\bibinfo{author}{\bibfnamefont{A.}~\bibnamefont{Wolf}},
  \bibinfo{author}{\bibfnamefont{J.~B.} \bibnamefont{Swift}},
  \bibinfo{author}{\bibfnamefont{H.~L.} \bibnamefont{Swinney}},
  \bibnamefont{and} \bibinfo{author}{\bibfnamefont{J.~A.}
  \bibnamefont{Vastano}}, \bibinfo{journal}{Physica D}
  \textbf{\bibinfo{volume}{16}}, \bibinfo{pages}{285} (\bibinfo{year}{1985}).

\bibitem[{\citenamefont{Zillmer et~al.}(2000)\citenamefont{Zillmer, Ahlers, and
  Pikovsky}}]{ZillmAhlersPik}
\bibinfo{author}{\bibfnamefont{R.}~\bibnamefont{Zillmer}},
  \bibinfo{author}{\bibfnamefont{V.}~\bibnamefont{Ahlers}}, \bibnamefont{and}
  \bibinfo{author}{\bibfnamefont{A.}~\bibnamefont{Pikovsky}},
  \bibinfo{journal}{Phys. Rev. E} \textbf{\bibinfo{volume}{61}},
  \bibinfo{pages}{332} (\bibinfo{year}{2000}).

\bibitem[{\citenamefont{Livorati et~al.}(2012)\citenamefont{Livorati, Kroetz,
  Dettmann, Caldas, and Leonel}}]{ibere12}
\bibinfo{author}{\bibfnamefont{A.~L.~P.} \bibnamefont{Livorati}},
  \bibinfo{author}{\bibfnamefont{T.}~\bibnamefont{Kroetz}},
  \bibinfo{author}{\bibfnamefont{C.~P.} \bibnamefont{Dettmann}},
  \bibinfo{author}{\bibfnamefont{I.~L.} \bibnamefont{Caldas}},
  \bibnamefont{and} \bibinfo{author}{\bibfnamefont{E.~D.}
  \bibnamefont{Leonel}}, \bibinfo{journal}{Phys.~Rev.~E}
  \textbf{\bibinfo{volume}{86}}, \bibinfo{pages}{036203}
  (\bibinfo{year}{2012}).

\bibitem[{\citenamefont{Contopoulos and Harsoula}(2010)}]{contopoulos10-1}
\bibinfo{author}{\bibfnamefont{G.}~\bibnamefont{Contopoulos}} \bibnamefont{and}
  \bibinfo{author}{\bibfnamefont{M.}~\bibnamefont{Harsoula}},
  \bibinfo{journal}{Celest.~Mech.~Dyn.~Astr.~} \textbf{\bibinfo{volume}{107}},
  \bibinfo{pages}{77} (\bibinfo{year}{2010}).

\bibitem[{\citenamefont{del Castillo-Negrete et~al.}(2005)\citenamefont{del
  Castillo-Negrete, Carreras, and Lynch}}]{negrete05}
\bibinfo{author}{\bibfnamefont{D.}~\bibnamefont{del Castillo-Negrete}},
  \bibinfo{author}{\bibfnamefont{B.~A.} \bibnamefont{Carreras}},
  \bibnamefont{and} \bibinfo{author}{\bibfnamefont{V.~E.} \bibnamefont{Lynch}},
  \bibinfo{journal}{Phys.~Rev.~Lett.~} \textbf{\bibinfo{volume}{94}},
  \bibinfo{pages}{065003} (\bibinfo{year}{2005}).

\bibitem[{\citenamefont{Gerlach et~al.}(2012)\citenamefont{Gerlach, W\"uster,
  and Rost}}]{rost12-1}
\bibinfo{author}{\bibfnamefont{M.}~\bibnamefont{Gerlach}},
  \bibinfo{author}{\bibfnamefont{S.}~\bibnamefont{W\"uster}}, \bibnamefont{and}
  \bibinfo{author}{\bibfnamefont{J.~M.} \bibnamefont{Rost}},
  \bibinfo{journal}{J.~Phys.~B} \textbf{\bibinfo{volume}{45}},
  \bibinfo{pages}{235204} (\bibinfo{year}{2012}).

\bibitem[{\citenamefont{Zagoya et~al.}(2012)\citenamefont{Zagoya, Goletz,
  Grossmann, and Rost}}]{rost12-2}
\bibinfo{author}{\bibfnamefont{C.}~\bibnamefont{Zagoya}},
  \bibinfo{author}{\bibfnamefont{C.~M.} \bibnamefont{Goletz}},
  \bibinfo{author}{\bibfnamefont{F.}~\bibnamefont{Grossmann}},
  \bibnamefont{and} \bibinfo{author}{\bibfnamefont{J.~M.} \bibnamefont{Rost}},
  \bibinfo{journal}{New J.~Phys.} \textbf{\bibinfo{volume}{14}},
  \bibinfo{pages}{093050} (\bibinfo{year}{2012}).

\end{thebibliography}

\end{document}